\documentclass[aps,prb,twocolumn,longbibliography,amsmath,amssymb,amsfonts,citeautoscript,floatfix]{revtex4-2}

\usepackage{graphicx}
\usepackage{bm}
\usepackage{float}
\usepackage{xcolor}
\usepackage{amsmath}
\usepackage{amssymb}
\usepackage[urlcolor=blue,colorlinks=true,citecolor=blue,linkcolor=blue,pdfstartview={FitH},bookmarks=false]{hyperref}

\newcommand{\ket}[1]{|#1\rangle}

\begin{document}

\title{Dynamical quantum phase transitions in a hybrid quantum dot system \\ with superconducting and ferromagnetic leads}

    \author{Antoni Jankiewicz}
    \email[e-mail:]{antjan1@amu.edu.pl}
    \affiliation{Institute of Spintronics and Quantum Information, Faculty of Physics and Astronomy, Adam Mickiewicz University, 61--614 Pozna\'n, Poland}
    
    \author{Kacper Wrze\'sniewski}
    \affiliation{Institute of Spintronics and Quantum Information, Faculty of Physics and Astronomy, Adam Mickiewicz University, 61--614 Pozna\'n, Poland}

    \author{Ireneusz Weymann}
    \affiliation{Institute of Spintronics and Quantum Information, Faculty of Physics and Astronomy, Adam Mickiewicz University, 61--614 Pozna\'n, Poland}

    \date{\today}


    \begin{abstract}	
        We theoretically explore the non-equilibrium dynamics of a single quantum dot system coupled to both ferromagnetic and superconducting electrodes. To investigate its time evolution, we utilize the time-dependent numerical renormalization group technique, which captures the system’s response to abrupt parameter changes in a fully non-perturbative manner.
        Our analysis focuses on dynamics following a sudden modification in the couplings to the leads or a shift of the orbital level. 
        In particular, we calculate the time evolution of the induced local superconducting pairing correlations and magnetization. In this context, the relevant energy spectra are examined.
	    Moreover, we study the behavior of the Loschmidt echo and the return function to shed light on the signatures of dynamical quantum phase transitions. The determined dependencies reveal non-trivial competition between relevant correlations, involving superconducting pairing and ferromagnetic-contacted induced exchange field, and deepen our understanding of nanoscale hybrid systems' dynamical behavior.
    \end{abstract}
    \maketitle

\section{Introduction}

    Hybrid mesoscopic devices that combine superconducting, ferromagnetic, semiconducting, and normal metallic elements have emerged as versatile platforms for exploring quantum phenomena driven by competition between fundamentally distinct orders. The theoretical description of such systems is commonly based on various Anderson-type impurity models~\cite{Anderson1961,Martin-Rodero2011,Balatsky2006May,Kennes2014Sep,Taranko2018Aug,Wrzesniewski2021Apr,Taranko2021Apr,Weymann2014Mar, Weymann2015Dec, Hwang2016Aug,Trocha2017Apr, Sonar2025Apr, Sonar2026Apr} that capture the essential physics of localized states coupled to reservoirs with different order parameters. The interplay between superconducting pairing, magnetic exchange, and Coulomb correlations gives rise to rich phase diagrams featuring quantum phase transitions between ground states of fundamentally different character.

    Within this theoretical landscape, hybrid quantum dot systems provide a versatile framework, serving both as highly controllable platforms for experimental implementations \cite{Hatter2015Nov,Hofstetter2010Jun, Heinrich2018Feb, GarciaCorral2020Mar, Huang2021Jul, Moehle2022Nov} and as models for sophisticated numerical simulations of non-equilibrium dynamics \cite{Anders2005,Kamp2021Jan,Kennes2014Sep, Taranko2018Aug,Wrzesniewski2021Apr,Taranko2021Apr, Antipov2016Jan, Wrzesniewski2022Mar, Slusarski2022Jun,Nghiem2017Oct,Nghiem2020Mar,Wrzesniewski2022,Bedow2022Nov,Ortmanns2023Aug,Wrzesniewski2019Jul}. Dynamical quantum phase transitions (DQPTs) have recently attracted growing attention, as they extend the concept of criticality into the time domain, offering a window into genuinely non-equilibrium aspects of quantum criticality inaccessible through static probes~\cite{Heyl2013, Heyl2018}. Understanding how systems traverse phase boundaries in real time is essential for advancing quantum technologies, where fast manipulation protocols inevitably drive devices far from equilibrium. Moreover, DQPTs provide a unifying framework connecting concepts from quantum information, many-body physics, and non-equilibrium thermodynamics, making them an important topic in modern condensed matter theory. 

    In this work, we present a theoretical and numerical investigation of dynamical quantum phase transitions in a hybrid mesoscopic structure consisting of a quantum dot sandwiched between ferromagnetic and superconducting leads, described by a generalized Anderson model. Employing the numerical renormalization group (NRG) method~\cite{Wilson1975, Bulla2008, Weichselbaum2012} and its time-dependent extension (td-NRG)~\cite{Anders2005, Anders2006, Nghiem2014Jul, Nghiem2018Oct, Nghiem2020Mar}, we analyze the non-equilibrium dynamics following quenches across different regions of the phase diagram and examine how the competition between interactions shapes the dynamical critical phenomena in these systems. In particular, we focus on
    examining the signatures of interplay between ferromagnetic-contact-induced exchange field
    and superconducting pairing correlations in the time-dependent expectation values
    of the on-dot pairing and magnetization. We demonstrate an oscillatory,
    counter-phase behavior of those quantities, which is
    revealed in the nonanalytic behavior of the return rate function,
    signaling the existence of dynamical quantum phase transitions in the system.

\section{Model and Hamiltonian}
    
\begin{figure}[h]
	        \includegraphics[width=1\linewidth]{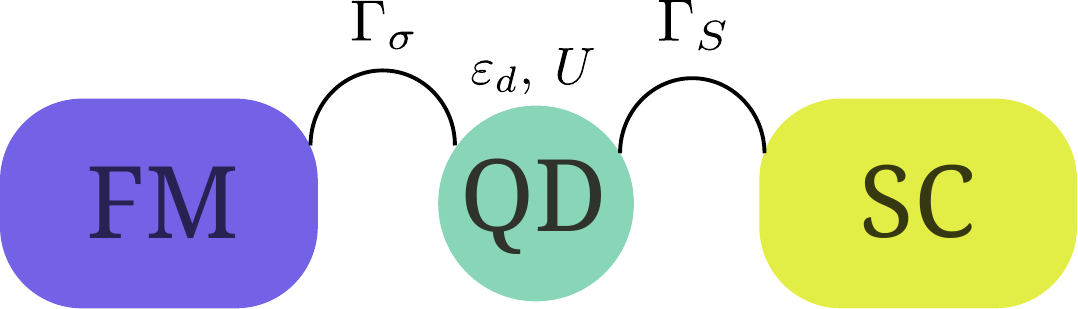}
            \caption{Schematic of a correlated quantum dot (QD) sandwiched
                between superconducting (SC) and ferromagnetic (FM) leads.
                The quantum dot hosts a spin-degenerate orbital level of energy $\varepsilon_d$
                where also Coulomb repulsion $U$ is present. The couplings to FM and SC leads
                are denoted as $\Gamma_\sigma$ (spin-dependent) and $\Gamma_S$, respectively.}
            \label{fig:model}
        \end{figure}

    This work focuses on a system containing a quantum dot coupled to both superconducting (SC) and ferromagnetic (FM) leads, schematically depicted in Fig.~\ref{fig:model}. 
        The Hamiltonian of the considered system is given by
        ${\hat{H}=\hat{H}_\text{QD}+\hat{H}_\text{FM}+\hat{H}_\text{FM-QD}+\hat{H}_\text{SC}}+\hat{H}_\mathrm{SC-QD}$, where
        \begin{equation}
            \hat{H}_\text{QD}=\sum_\sigma \varepsilon_d \hat{d}^\dagger_\sigma \hat{d}_\sigma + U \hat{d}^\dagger_\uparrow \hat{d}_\uparrow \hat{d}^\dagger_\downarrow \hat{d}_\downarrow
        \end{equation}
        represents the quantum dot; $\hat{d}^\dagger_\sigma$($\hat{d}_\sigma$) creates (annihilates) an electron of spin $\sigma$ that occupies the dot;
        $\varepsilon_d$ accounts for the dot's energy level and $U$ is responsible for the Coulomb interactions.

        The Hamiltonian of the ferromagnetic lead $\hat{H}_\text{FM}$ is expressed as
        \begin{equation}
            \hat{H}_\text{FM}=\sum_{k\sigma} \varepsilon_{k\sigma} \hat{c}^\dagger_{k \sigma} \hat{c}_{k\sigma}
        \end{equation}
        where $\hat{c}_{k \sigma}^\dagger$($\hat{c}_{k\sigma}$)
        is the creation (annihilation) operator of an electron
        with momentum $k$, spin $\sigma$ and energy $\varepsilon_{k\sigma}$.

        The next term of the total Hamiltonian, $\hat{H}_\text{FM-QD}$, 
        describes the tunneling of electrons with momentum $k$ and
        spin $\sigma$ from the ferromagnetic electrode to the quantum dot, as well as the reverse process

        \begin{equation}
            \hat{H}_\text{FM-QD}=\sum_{k\sigma} V_{k \sigma} (\hat{d}^\dagger_\sigma \hat{c}_{k\sigma} + \hat{c}^\dagger_{k\sigma}\hat{d}_\sigma),
        \end{equation}
        with $V_{k \sigma}$ being the relevant tunnel matrix element.
        The hybridization with the FM lead gives rise 
        to the broadening of the quantum dot level,
        which can be expressed as $\Gamma_\sigma \equiv \pi\rho_\sigma |V_\sigma|^2$,
        where $\rho_\sigma$ is the spin-dependent density of states 
        of the normal lead and we assume momentum-independent 
        tunnel matrix elements ${V_{\sigma} \equiv  V_{k\sigma}}$.
        The coupling can be further written in terms
        of spin-polarization of the lead $p$ as
        $\Gamma_\sigma = (1+\sigma p)\Gamma$,
        where ${\sigma=+}$ for spin-up and ${\sigma=-}$ for spin-down electrons,
        while ${\Gamma \equiv (\Gamma_\uparrow + \Gamma_\downarrow)/2}$.

        Finally, $\hat{H}_\text{SC}$
        describes the superconductor and $\hat{H}_\mathrm{SC-QD}$ represents the coupling
        to the quantum dot. In the following we employ the superconducting atomic
        limit to account for both terms in our calculations.
        This approximation captures the low-energy physics by
        integrating out the high-energy degrees of freedom associated
        with the superconducting lead, which results in the following effective Hamiltonian \cite{Rozhkov2000,Meng2009,Tanaka2007,Vecino2003}
        \begin{equation}
            \hat{H}_\text{SC}+\hat{H}_\mathrm{SC-QD}\approx\Gamma_S (\hat{d}^\dagger_\uparrow \hat{d}^\dagger_\downarrow+\hat{d}_\downarrow \hat{d}_\uparrow),
        \end{equation}
        where the presence of the SC lead is manifested through
        a local pairing induced in the quantum dot and characterized by the Andreev reflection rate $\Gamma_S$.

	\section{Method and quantities of interest}
    
        Our calculations are based on the numerical renormalization group method~\cite{Wilson1975, Bulla2008}, which allows us to accurately capture the low-energy physics and identify static properties of the system. The core of NRG is a logarithmic discretization of the conduction band and mapping of such a discretized Hamiltonian on a one-dimensional, semi-infinite chain with decaying hopping integrals. Separation of the energy scale between consecutive
        chain sites enables efficient iterative calculations by keeping
        a fixed number of low-energy states along the chain \cite{Bulla2008}. 
        
        The static NRG method forms the framework of the time-dependent numerical renormalization group method~\cite{Anders2005,Anders2006,Wrzesniewski2019Jul,Weichselbaum2012}, which, on the other hand, enables us to evaluate the dynamics that follows a quantum quench, i.e.\ to examine the time-dependence of relevant observables and overlaps. 
        In td-NRG, one performs a diagonalization of both the initial and final Hamiltonians, the eigenbases of which are then used to determine the relevant dynamical properties.
        One of the central quantities describing the dynamical behavior of the system and allowing for the identification of DQPTs is the Loschmidt echo~\cite{Goussev2012} that measures how the system departs from its ground state
        \begin{equation}
            \mathcal{L}(t)\equiv \bigl|\langle \Psi_0 | \Psi(t) \rangle\bigl|^2 =\bigl|\langle\Psi_0|e^{-i\hat{H}t}|\Psi_0\rangle\bigl|^2.
        \end{equation}
        Here, $|\Psi(t)\rangle$ denotes the state of the system after a quench at time $t$, $\langle \Psi_0|$ is a dual vector to the one that describes the ground state of the system in the initial conditions ($t=0$), while $\hat{H}$ denotes the Hamiltonian of the system after the quench.

        It is interesting to note a similarity of the Loschmidt echo to the partition function from the theory of statistical physics and classical phase transitions \cite{Heyl2013,Heyl2018}. In analogy with this, one can define a return function as a counterpart of the Helmholtz free energy
        \begin{equation}
            \lambda(t)=- \lim_{N\to\infty}\frac{1}{N} \ln \mathcal{L}(t),
        \end{equation}
        where $N$ denotes the number of degrees of freedom.
        By analogy to classical phase transitions, dynamical quantum phase transitions are defined at times for which $\lambda(t)$ exhibits non-analytic behavior \cite{Heyl2013}. This usually happens when the quench is performed between two distinct quantum phases of the system \cite{Heyl2018}.

\begin{figure}[t]
\centering
    \includegraphics[width=1\linewidth]{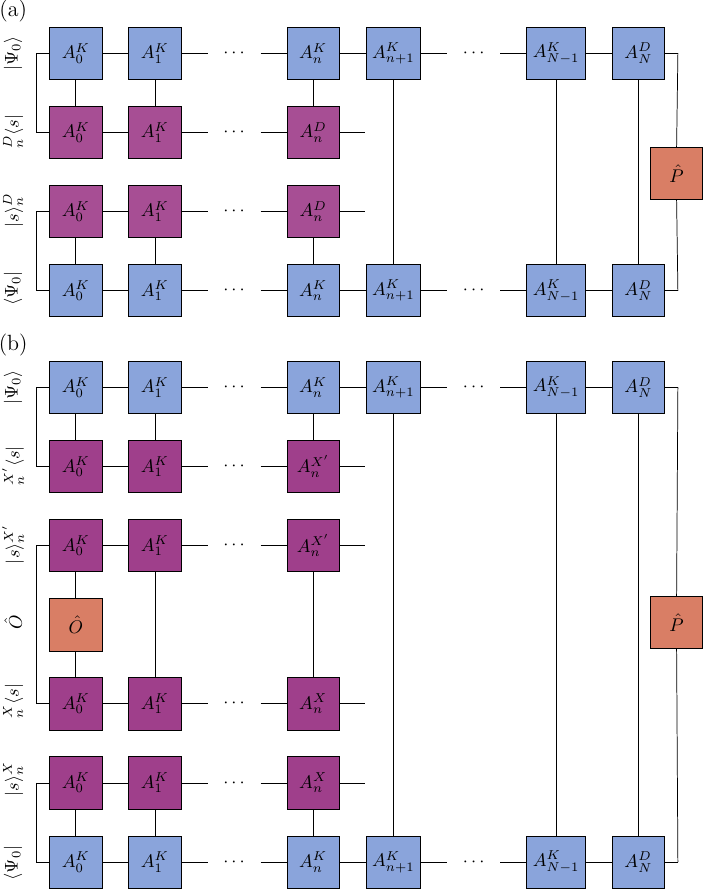}
    \caption{Graphical representation of the matrix product state diagrams
    for the computation of the coefficients (a) $C_{ns}$ and 
    (b) $\Theta_{nss'}^{XX'}$.
    Blue blocks represent the MPS tensors belonging to the initial Hamiltonian $\hat{H}_0$,
    while purple blocks denote tensors of the final Hamiltonian $\hat{H}$.
    $\hat{P}$ is a projector onto the ground state of $\hat{H}_0$.}
    \label{Fig:MPS_C}
\end{figure}

To facilitate computations of the time-dependent quantities, 
we perform the calculations within the tensor network formalism.
In this framework, the quantum states of the system defined on the chain can be efficiently represented as matrix product states (MPS)~\cite{Weichselbaum2012}, given by
		\begin{equation}
            |s\rangle_n^X \equiv \sum_{s,\sigma} A^{X[\sigma_n]}_{s_{n-1},s_n} |\sigma\rangle_n |s\rangle_{n-1}^K, 
        \end{equation}
where $\ket{s}_n^X$ denotes a state kept ($X=K$) or discarded ($X=D$) at the chain site $n$, $A^{X[\sigma_n]}_{s_{n-1},s_n}$ is the corresponding coefficient tensor and $\ket{\sigma}_n$ is a local state. The above states, supplemented by environmental states $\ket{e}$ that describe the rest of the chain, form the full many-body eigenbasis of the Hamiltonian~\cite{Anders2005}
		\begin{equation} 
            \hat{\mathcal{I}} = \sum_{n,s,e} |se\rangle_n^D {}_n^D\langle se|.
        \end{equation}
This eigenbasis, obtained for both the initial and final Hamiltonians, is then used to determine the Loschmidt amplitude ${\mathcal{G}(t) \equiv \langle \Psi_0 | \Psi(t) \rangle }$, which is given by
\begin{equation}
\begin{gathered}
\mathcal{G}(t) =
\langle \Psi_0 | e^{-i\hat{H}t} | \Psi_0 \rangle\\
= \sum_{n,s,e} \langle \Psi_0 | e^{-i\hat{H}t} | se \rangle_n^D {}_n^D\langle se | \Psi_0 \rangle \\
= \sum_{n,s,e} e^{-i E_{ns}^D t}
\left| \langle \Psi_0 | se \rangle_n^D \right|^2\\
=\sum_{n,s}d^{N-n}e^{-iE_{ns}^Dt}|
\langle\Psi_0|s\rangle_n^D|^2\\
\equiv \! \sum_{n,s} d^{N-n} \, e^{-i E_{ns}^D t}\,C_{ns},
\end{gathered}
\end{equation}
where $E_{ns}^D$ denotes the eigenenergy of the state $\ket{se}_n^D$ and $d$ stands for state space dimension of single Wilson chain site.
A matrix product state representation of a diagram for the computation of the coefficients $C_{ns}$ is presented in Fig.~\ref{Fig:MPS_C}(a).

Similarly, a time-dependent expectation value of an operator of interest
$\hat{\mathcal{O}}(t)$ can be found from the following formula
\begin{widetext}
        \begin{equation}\label{eq:observable}
            \begin{gathered}
		\langle \hat{O} \rangle (t)=\langle \Psi_0 |e^{i\hat{H}t}\hat{O} e^{-i\hat{H}t}|\Psi_0\rangle=
		\sum_{n,n'}\sum_{s,s',e,e'} \langle \Psi_0 |e^{i\hat{H}t}|s e \rangle_n^D {}_n^D\langle s e| \hat{O} e^{-i\hat{H}t} |s' e' \rangle_{n'}^D {}_{n'}^D\langle s' e'|\Psi_0\rangle\\
		=\sum_{n,n'}\sum_{s,s',e,e'} e^{i (E_{n s}^D-E_{n' s'}^{D})t}\delta_{e,e'}\langle \Psi_0 |s e \rangle_n^D {}_n^D\langle s| \hat{O} |s' \rangle_{n'}^D {}_{n'}^D\langle s' e'|\Psi_0\rangle \\
		=\sum_{n}^{X,X'\ne K,K} \sum_{s,s'} d^{N-n}e^{i (E_{n s}^X-E_{n s'}^{X'})t}\langle \Psi_0 |s  \rangle_n^X {}_n^X\langle s | \hat{O} |s' \rangle_{n}^{X'} {}_{n}^{X'}\langle s' |\Psi_0\rangle\equiv\sum_{n}^{X,X'\ne K,K} \sum_{s,s'}d^{N-n} e^{i (E_{n s}^X-E_{n s'}^{X'})t}\Theta_{nss'}^{XX'},
\end{gathered}
        \end{equation}
\end{widetext}
where the corresponding MPS diagram for the calculation of $\Theta_{nss'}^{XX'}$
is presented in Fig.~\ref{Fig:MPS_C}(b).

The relevant expectation values and Loschmidt echo are calculated in the frequency domain \cite{Weichselbaum2012},
in a similar fashion to spectral functions in static NRG calculations,
and then Fourier-transformed to the time domain. The transform of the Loschmidt amplitude can be expressed as
\begin{equation}
        \mathcal{G}(\omega)=\sqrt{2\pi}\sum_{n,s}d^{N-n}C_{ns}\delta(\omega+E_{ns}^D),
\end{equation}
with $\delta$ denoting the Dirac's delta function.
The transform for the expectation values of the observables is given by
\begin{equation}
            \langle\hat{\mathcal{O}}\rangle(\omega)=\sqrt{2\pi}\sum_n^{X,X'\ne K,K}\sum_{s,s'}d^{N-n}\Theta_{nss'}^{XX'}\delta(\omega+E_{n,s'}^{X'}-E_{ns}^X).
\end{equation}
The analytical formulations derived within both the temporal and spectral domains serve to elucidate the complex underlying dynamics of the system, providing a holistic perspective that is subjected to a rigorous and systematic investigation in Section IV B.

\section{Numerical results and discussion}

In this section, we present and discuss the numerical results
on the considered hybrid quantum dot system.
We first analyze its static behavior and then focus on examining the
dynamical properties; in both cases we examine the behavior of
the z-th projection of the local spin operator $\hat{S}_z\equiv \frac{1}{2}(\hat{d}^\dagger_\uparrow \hat{d}_\uparrow-\hat{d}^\dagger_\downarrow \hat{d}_\downarrow$)
and the local pairing operator $\hat{d}_\downarrow \hat{d}_\uparrow$.

All presented results were obtained for the following parameters:
$U=0.1$ (in units of band halfwidth $D\equiv1$),
$\Gamma=0.02U$ and the spin polarization of ferromagnetic lead $p=0.5$.
We also consider the zero-temperature case $T=0$.
Moreover, to perform the numerical renormalization group calculations,
we assume the discretization parameter $\Lambda=2.2$
and keep $N_\text{kept}=2000$ states during iterative diagonalization.
We also make use of the Abelian spin symmetry \cite{FlexibleDMNRG,Wrzesniewski2019Jul}.

\subsection{Static properties}

\begin{figure}[h]
    \includegraphics[width=\linewidth]{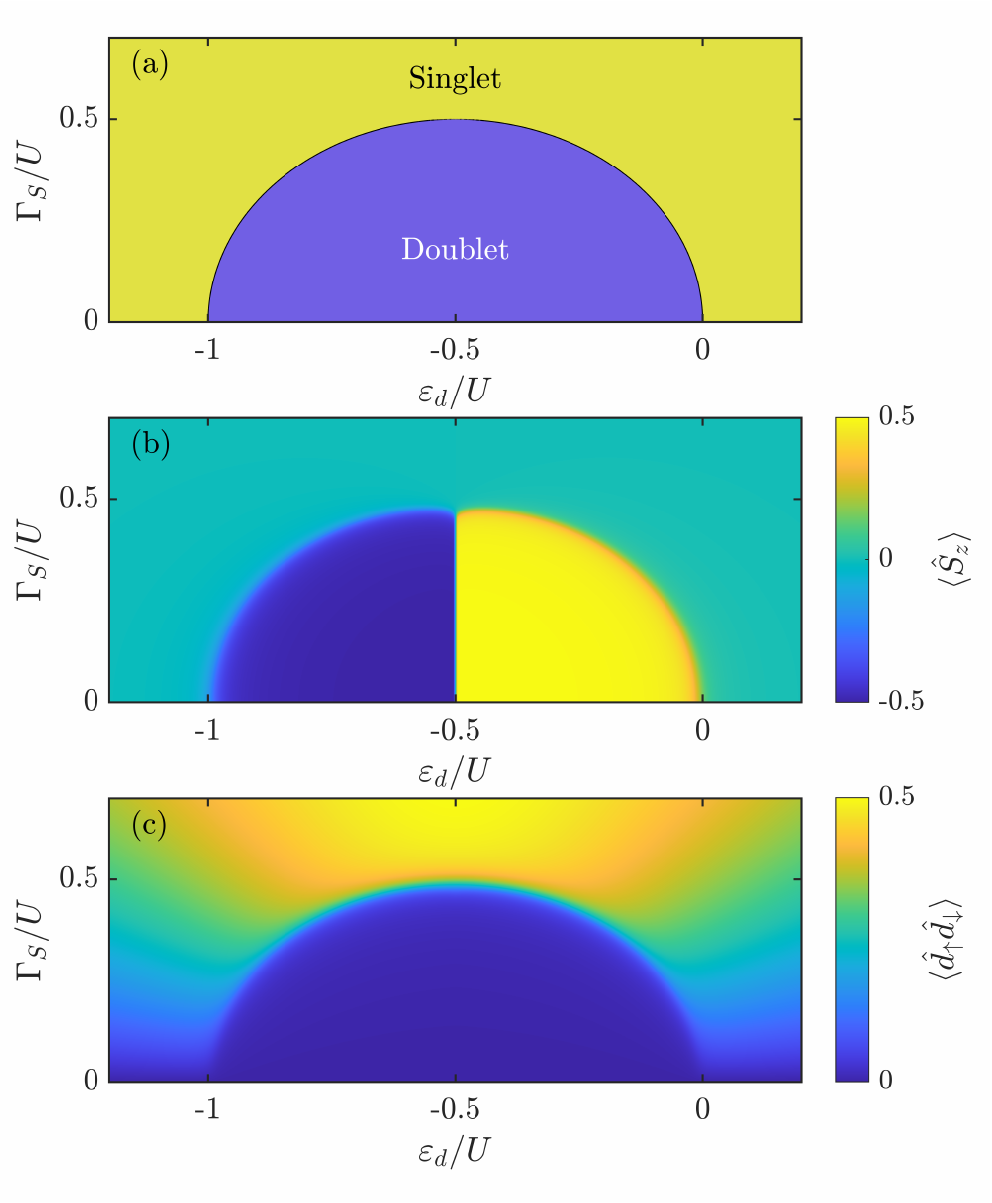}
        \caption{The static properties of the considered hybrid system:
        (a) the phase diagram,
        (b) the spin expectation value,
        and (c) the superconducting pairing
        calculated as a function of the dot's energy level $\varepsilon_d$
        and superconducting pairing amplitude $\Gamma_S$.}
    \label{fig:phase_diagram}
\end{figure}

To facilitate the discussion of dynamical properties, let us first analyze the static behavior of the system.
A simple analysis of the system uncoupled from the ferromagnetic lead ($\Gamma=0$) yields two phases:
a spin-doublet characterized by single-electron states $|{\uparrow}\rangle$ and $|{\downarrow}\rangle$
and spin-singlet characterized by BCS-type ground state ${|\pm\rangle=\alpha_\mp|0\rangle\pm\alpha_\pm|{\uparrow\downarrow}\rangle}$,
where probability amplitudes are defined as $\alpha_\pm\equiv\frac{1}{\sqrt{2}}\sqrt{1\pm\frac{\delta}{\sqrt{\delta^2+\Gamma_S^2}}}$
and $\delta\equiv\frac{U}{2}+\varepsilon_d$ denotes detuning from the particle-hole symmetry point ($\varepsilon_d = -U/2$).
The corresponding phase diagram is shown in Fig.~\ref{fig:phase_diagram}(a).
The phase boundary between the singlet and doublet phases 
is given by \cite{Bauer2007}
\begin{equation}
    \Gamma_S^2+\biggl(\varepsilon_d+\frac{U}{2}\biggl)^2=\biggl(\frac{U}{2}\biggl)^2.
\end{equation}
In the case of ferromagnetic lead, although the phase boundary 
is not affected by the spin polarization and thus resembles its non-magnetic counterpart \cite{Bauer2007}, the symmetry of the doublet phase is reduced.
In particular, the spin-resolved charge fluctuations 
between the quantum dot and FM contact result in an exchange field \cite{Martinek2003Dec,MartinekPRL2003,PasupathyScience2004,MartinekPRB2005,GaassPRL2011}. In consequence, the doublet state becomes split and
the quantum dot exhibits finite magnetization, see Fig.~\ref{fig:phase_diagram}(b), the sign of which changes when one crosses the particle-hole symmetry point.
The magnitude of the exchange field for our hybrid quantum dot device
can be found within the second-order perturbation theory, which gives
the following expression \cite{Wojcik2014Apr}
\begin{equation}\label{eq:exchange}
    \varepsilon_\text{ex}=\frac{2p\Gamma}{\pi}\frac{\delta}{\sqrt{\delta^2+\Gamma_S^2}}\bigg[\phi(E_-)-\phi(E_+)\bigg],
\end{equation}
where $\phi(\omega)\equiv\Re[\Psi(\frac{1}{2}+i\frac{\omega}{2\pi T})]$, and $\Psi(\omega)$ denotes the digamma function, whereas 
\begin{equation}\label{eq:andreevenergies}
    E_\pm\equiv\frac{U}{2}\pm\sqrt{\delta^2+\Gamma_S^2}
\end{equation}
stand for the corresponding Andreev excitation energies.

On the other hand, the expectation value of the superconducting pairing $\langle \hat d_\downarrow \hat d_\uparrow \rangle$, presented in 
Fig.~\ref{fig:phase_diagram}(c), exhibits finite value in the singlet phase,
with increasing magnitude as one moves to higher values of the pairing 
amplitude $\Gamma_S$.

\subsection{Dynamical properties}

    We have performed the td-NRG calculations of quantum quenches
    in the coupling to the superconducting electrode $\Gamma_S$,
    as well as in the energy level of the quantum dot $\varepsilon_d$. 
    First of all, we note that for quenches from the doublet to the singlet phase,
    the qualitative dynamical behavior weakly depends
    on the choice of the initial value of $\Gamma_S$.
    Therefore, for all quenches in the coupling to the superconductor,
    the initial coupling to the superconducting lead is assumed to be $\Gamma_{S0}=0$.    
    On the other hand, in the case of quenches in the dot's energy level,
    we assume $\varepsilon_{d0}=-0.4U$, which is close to the electron-hole
    symmetry point but with detuning $\delta$ sufficient to split the doublet state,
    cf. the formula for the exchange field (\ref{eq:exchange}).

    \begin{figure}[t]
        \includegraphics[width=1\columnwidth]{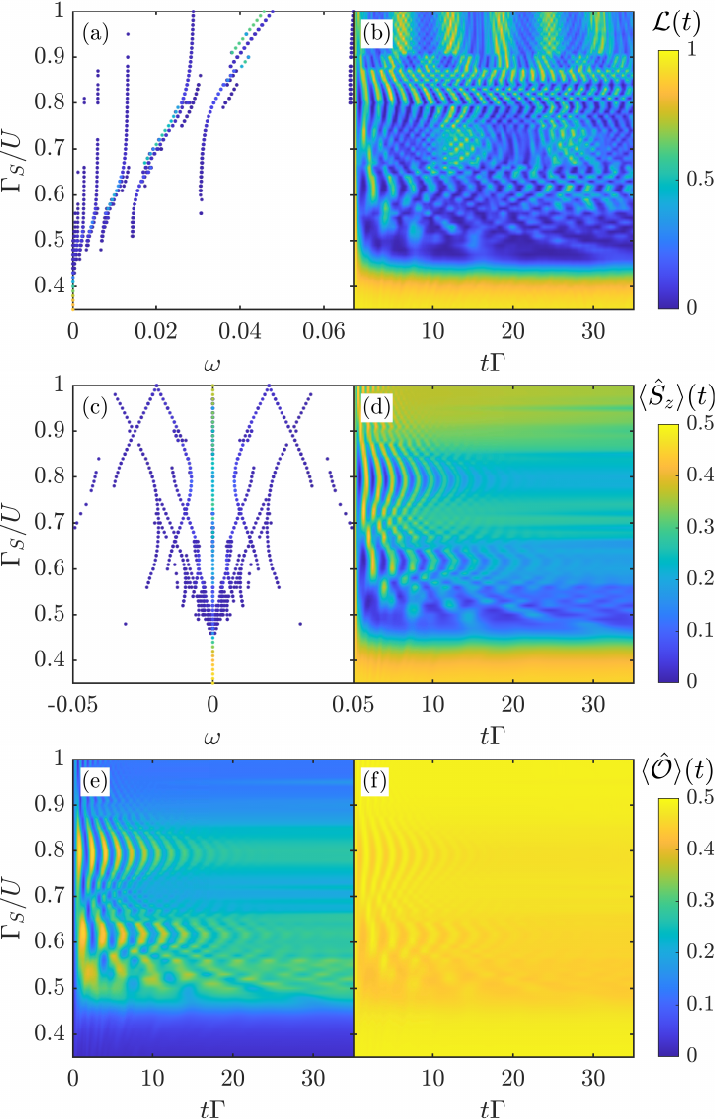}
        \caption{(b),(d),(e),(f) plots illustrating the time dependence of (b) Loschmidt echo, (d) local magnetization, (e) local pairing ($\langle\hat{\mathcal{O}}\rangle(t)=\Re[\langle \hat{d}_\downarrow \hat{d}_\uparrow\rangle(t)]$) and (f) half of the occupation number ($\langle\hat{\mathcal{O}}\rangle(t)=\langle\hat{n}\rangle(t)/2$) after a quench in the coupling to the superconductor $\Gamma_S$. The x-axis represents time, the y-axis represents the final coupling strength to the superconducting lead $\Gamma_S$, and the color scale indicates expectation value of the relevant observable. The corresponding frequency spectra (collected delta peaks) are shown in (a), (c) where the x-axis represents the oscillation frequency. The y-axis and color bar are shared among adjacent plots. The color of the dots in the frequency spectra indicates magnitudes of delta peaks contributions.
        The calculations were performed for $\varepsilon_d=-0.4U$.}
        \label{fig:2dquenchgs}
    \end{figure}

        \begin{figure}[t]
        \includegraphics[width=1\columnwidth]{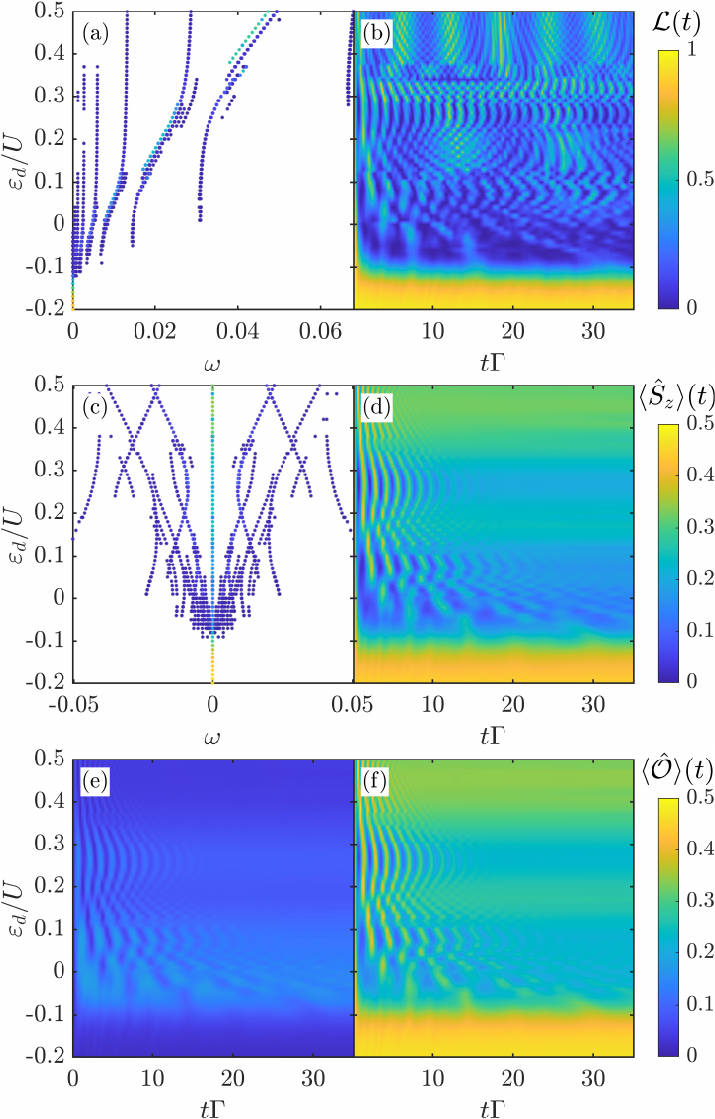}
            \caption{(b),(d),(e),(f) plots illustrating the time dependence of (b) Loschmidt echo, (d) local magnetization, (e) local pairing ($\langle\hat{\mathcal{O}}\rangle(t)=\Re[\langle \hat{d}_\downarrow \hat{d}_\uparrow\rangle(t)]$) and (f) half of the occupation number ($\langle\hat{\mathcal{O}}\rangle(t)=\langle\hat{n}\rangle(t)/2$) after a quench. The x-axis represents time, the y-axis represents the final coupling strength to the superconducting lead $\varepsilon_d$, and the color scale indicates expectation value of the relevant observable. The corresponding frequency spectra (collected delta peaks) are shown in (a), (c) where the x-axis represents the oscillation frequency. The y-axis and color bar are shared among adjacent plots. The color of the dots in the frequency spectra indicates magnitudes of delta peaks contributions. The calculations were performed for $\Gamma_S/U=0.25$.}
        \label{fig:2dquenchen}
    \end{figure}

    Firstly, we study a quench in the dot–superconductor coupling $\Gamma_S$, where the system is prepared in the ground state of the initial Hamiltonian $\hat{H_0}$ with $\Gamma_S=0$ and suddenly coupled to the superconducting lead with a finite final value $\Gamma_S/U$, indicated on the vertical axis of all six panels of Fig.~\ref{fig:2dquenchgs}. The horizontal axis of the panels (b), (d), (e) and (f) is the time elapsed after the quench, $t \Gamma$, so that scanning along a fixed horizontal line shows the subsequent time evolution of (b) the Loschmidt echo $\mathcal{L}(t)$, (d) the local magnetization $\langle \hat{S}_z(t) \rangle$, (e) the local pairing $\Re[\langle \hat{d}_\downarrow \hat{d}_\uparrow\rangle(t)]$, and (f) half of the dot's occupation $\langle \hat{n}\rangle/2$ for that particular quench amplitude. The top two left-hand panels [Fig.~\ref{fig:2dquenchgs}(a) and (c)] represent the discrete frequency data of the dynamics for the same set of final $\Gamma_S/U$ values. Each point marks the frequency $\omega_n$ of a delta peak that contributes to the corresponding observable, with color encoding its spectral weight. These frequencies correspond to energy differences between many-body excited states of the final Hamiltonian $\hat H$ that are reached from the initial state, and it is the interference of these discrete modes, upon the Fourier transform, that generates the time traces shown on the right. Also, the behavior of local pairing and occupation number is similar to that of spin, so we omit showing the corresponding spectra. Notably, the oscillations of the superconducting pairing are in counterphase to those of the spin expectation value. We will show this behavior below in the corresponding cross-section plots.

The structures visible in Fig.~\ref{fig:2dquenchgs}(a) and (c) directly explain the qualitative features of the time-dependent panels (b) and (d). Quenches that leave the system in the initial phase cause the expectation values to relax rapidly on a time scale set by $\Gamma$ from the values given by the average with respect to $\hat H_0$ to those of $\hat H$ [as can be seen in the lower parts of Fig.~\ref{fig:2dquenchgs}(d),(e) and (f)], accompanied by a drop in the Loschmidt echo [Fig.~\ref{fig:2dquenchgs}(b)]. The magnitude of the drop is proportional to the difference between these expectation values. The spectra of these dynamics are dominated by a single zero frequency mode with the magnitude of the relevant steady-state expectation value with respect to $\hat H$. 
For a small range of parameters right above the phase boundary one can observe fast relaxation to the spin-singlet phase, as indicated by the disappearance of dot's spin expectation value, growth in pairing and finally a dynamical quantum phase transition as echo approaches zero. The overall increase in the number of resolved branches and their spread in frequency as $\Gamma_S/U$ grows [see Fig.~\ref{fig:2dquenchgs}(c)] tracks the increasing complexity of the low-lying many-body spectrum of $\hat H$ as the effective pairing becomes stronger.

For higher values of the parameters, we can divide our analysis into two distinct parameter regions. In the first one, around $\Gamma_S\approx0.81U$, $\Gamma_S\approx 0.62U$, the dot magnetization oscillates even
twice longer than in the neighboring interference-dominated regions (around $\Gamma_S\approx0.72U$ and $\Gamma_S\approx1U$), and relaxes to significantly lower values. Moreover, the oscillation damping rate is higher for larger values of the quenched parameter $\Gamma_S$. The spectrum of these oscillations exhibits slightly weaker modes at non-zero frequencies. The non-zero frequency branches show frequency minima in the center of the region. As can be seen in the figure, the dynamics of the magnetization is accompanied by well behaving oscillations of the Loschmidt echo,
which dominant frequency matches the one of the magnetization. Furthermore,
in the regions' centers, the Loschmidt echo's spectrum
displays anticrossing features.

On the other hand, in the region where the spin's oscillations are dominated by the interference patterns ($\Gamma\approx 0.58U$, $\Gamma_S\approx0.72U$, $\Gamma_S\approx1U$), oscillations tend to settle at higher values of $\langle \hat{S}_z \rangle$. Their spectra are dominated by higher magnitude zero frequency modes accompanied by low frequency modes ($\omega\approx0$). In addition, several branches of comparable weight run close to one another.
In the case of Loschmidt echo, a complex character is exhibited, where a long-wave pattern weakly dependent on $\Gamma_S$ interferes with multiple lower frequencies, see the first row of Fig.~\ref{fig:2dquenchgs}.

A quench in the level position $\varepsilon_d$, when compared to a quench in $\Gamma_S$, produces qualitatively very similar dynamics (see Fig.~\ref{fig:2dquenchen}), which is a direct consequence
of how such a quench reshapes the Andreev bound-state
energies discussed above Eq.~(\ref{eq:andreevenergies}). Thereby,
the quench tunes the Andreev splitting in essentially
the same way as changing $\Gamma_S$ at fixed $\varepsilon_d$ does. Consequently, the same set of low-lying many-body transitions is excited by the quench, and the resulting time dependencies of $\mathcal{L}(t)$
and $\langle \hat{S}_z(t) \rangle$ show similar characteristic oscillations
and beating patterns as for the quench in $\Gamma_S$,
controlled by the same underlying energy scale $\sqrt{\delta^2+\Gamma_S^2}$.
However, an additional feature specific to the quench in $\varepsilon_d$
is that it also drives charge dynamics in the quantum dot
[see Fig.~\ref{fig:2dquenchen}(f)], while suppressing local pairing [Fig.~\ref{fig:2dquenchen}(e)]. This is because $\varepsilon_d$
couples directly to the dot occupation rather than only to the pairing amplitude. We note that the charge oscillations and their relation to the Andreev-state structure has already been analyzed
in the work on the nonmagnetic \cite{Wrzesniewski2021Apr}, where qualitatively similar oscillatory behavior was reported.
        
    \begin{figure}[t]
        \includegraphics[width=1\linewidth]{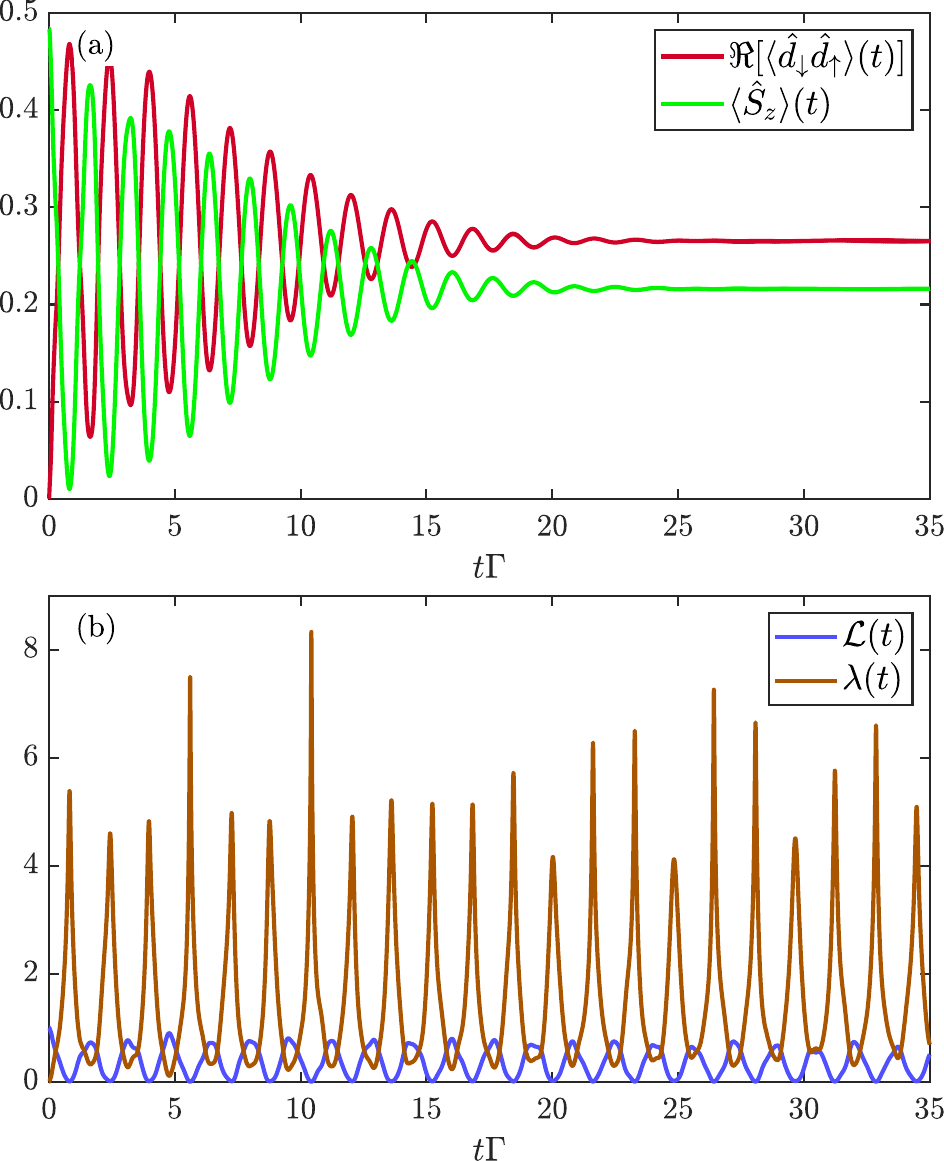}
        \caption{Time-dependence of: (a) real part of local superconducting pairing (red), local spin (green) and (b) Loschmidt echo (blue) along with return rate (brown). The results were obtained for the following parameters: $\: \varepsilon_d=-0.4U,\: \Gamma_{S0}=0\to\Gamma_S=0.81U$.}
        \label{fig:crosssection_phtr}
    \end{figure}

In Fig.~\ref{fig:crosssection_phtr}(a) we show the time evolution of two complementary quantities: the real part of the average pairing amplitude $\Re\bigl[\langle \hat d_\downarrow \hat d_\uparrow\rangle(t)\bigl]$ and the local magnetization $\langle \hat{S}_z\rangle(t)$. Both quantities oscillate with the same frequency, since they originate from the same pairs of many-body states in the excitation spectrum of $\hat H$, but in {\em counter-phase}. Evidently, as the pairing correlation increases, the magnetization decreases, and vice versa, showing mutually exclusive competition driven by these two orderings. 

This reflects the competition between two ordering mechanisms available to the dot electrons: a spin-singlet BCS-like pairing that correlates
$\hat d_\downarrow\hat d_\uparrow$ 
and a spin-polarized configuration favored by the exchange field induced by the ferromagnetic electrode.
The envelope of both oscillations decays approximately exponentially in time $\sim e^{-t/\tau}$ with the timescale given by $\tau \propto 1/\Gamma$. Here, the coupling to the leads gives the excited many-body states a finite lifetime, leading to dephasing of the Rabi-like oscillations between them.
After $t\Gamma \gtrsim 20$ both quantities reach stationary asymptotic values, corresponding to the diagonal ensemble in the post-quench Hamiltonian $\hat H$,
see Fig.~\ref{fig:crosssection_phtr}(a).
The Loschmidt echo together with the corresponding return rate $\lambda(t)$
is shown in Fig.~\ref{fig:crosssection_phtr}(b).
The echo oscillates strongly and periodically approaches zero values — precisely at those instants when the time-evolving state $|\psi(t)\rangle$ becomes orthogonal to the initial state $|\psi(0)\rangle$. Such events correspond to sharp, narrow divergences in $\lambda(t)$, commonly referred to as kinks. These non-analyticities of the return-rate function in time, in the context of dynamical quantum phase transitions, are interpreted as the real-time analogues of equilibrium phase transitions. Non-analytic behavior signals abrupt changes in the structure of the overlap between the initial state and the eigenstates of $\hat H$ as the system time evolves, indicating dynamical quantum phase transitions.
    
\begin{figure}[h]
        \includegraphics[width=1\linewidth]{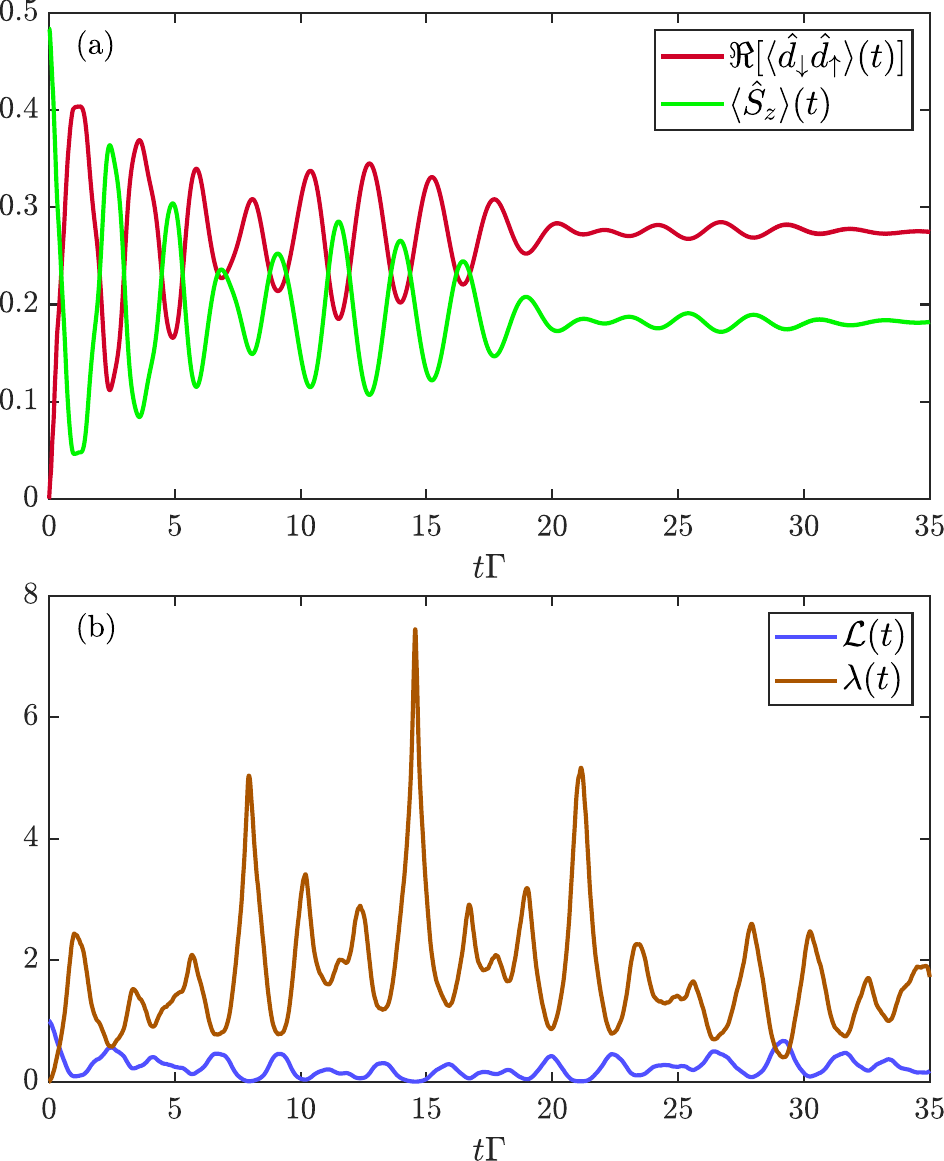}
        \caption{Time-dependence of: (a) real part of local superconducting pairing (red), local spin (green) and (b) Loschmidt echo (blue) along with return rate (brown). The results were obtained for the following parameters: $\: \varepsilon_d=-0.4U,\: \Gamma_{S0}=0\to\Gamma_S=0.59U$.}
        \label{fig:crosssection_beat}
    \end{figure}

In the case of the region dominated by interferences, which is presented in  Fig.~\ref{fig:crosssection_beat}, pronounced beating patterns and a faster decay of the oscillation amplitude can be observed.       
The irregular, quasiperiodic occurrence of the kinks
(rather than strictly periodic) in the Loschmidt echo indicates a multimode character of the dynamics. Several competing frequencies $\omega_n$ are present in the spectrum discussed previously, rather than a single dominant oscillation frequency.

An essential advantage of hybrid quantum-dot-based realizations of setups exhibiting critical behavior in time is a high degree of tunability. All gate-controlled parameters, such as $\varepsilon_d$, $\Gamma_S$, and the exchange splitting from the ferromagnetic lead can be adjusted essentially independently and continuously. This makes it possible to reliably tune the system to an appropriate point in parameter space where a single frequency behavior dominates the post-quench dynamics rather than the generic multimode beating discussed above. In such a regime, the time evolution of relevant observables reduces to essentially single-mode, well-resolved oscillations whose extrema directly correlate with the occurrence of kinks in the return rate $\lambda(t)$, signaling the presence of dynamical quantum transitions.
Crucially, in the present hybrid system, the local magnetization emerges as a particularly pivotal observable that directly tracks such non-equilibrium dynamics. Beyond serving as a theoretical diagnostic, $\langle \hat{S}_z(t)\rangle$ could in principle be experimentally detected via time-resolved spin-transport or spin-detection techniques. This highlights the strong potential of local spin dynamics as a sensitive experimental probe for identifying dynamical quantum phase transitions in controlled,
tunable solid-state devices.

	\section{Summary}

We have investigated the dynamical behavior of a hybrid quantum dot system
consisting of a quantum dot coupled to superconducting and ferromagnetic leads.
We have revealed a distinct phase diagram of the system characterized by 
transitions between singlet and doublet ground states, determined by the interplay of quantum dot energy level position and the coupling strength to the superconductor $\Gamma_S$. We have shown that, following a quantum quench, the time evolution of local observables, specifically the local magnetization $\langle \hat S_z \rangle$ and the local pairing amplitude $\langle \hat d_\downarrow \hat d_\uparrow\rangle$, exhibit complex oscillatory patterns and relaxation processes. In particular, these quantities were shown to oscillate in counter-phase with respect to each other, which is a direct consequence of the competition between magnetic and superconducting correlations.
Finite dot magnetization results from the presence of an exchange field stemming from the ferromagnetic lead, which induces a splitting in the doublet state that is reflected in the characteristic frequency components of the magnetization's power spectrum. 

Moreover, by analyzing in detail the dynamical behavior of the Loschmidt echo $\mathcal{L}(t)$ and the corresponding return rate $\lambda(t)$, we have identified nonanalytic kinks in $\lambda(t)$, which constitute a hallmark of dynamical quantum phase transitions. 
Such behavior serves as a definitive signature of the dynamical quantum phase transitions, a phenomenon that we have further corroborated by the analysis of the frequency spectra
obtained from the delta-peak distributions.
We have observed dynamical quantum phase transitions for quenches
between two distinct phases of the system, accompanied by characteristic signatures in the dynamics of local observables, such as dot magnetization and on-dot pairing. Our results highlight how the competition between superconducting pairing
and ferromagnetic exchange effects fundamentally shapes 
the complex dynamical landscape of the system.

\begin{acknowledgements}
This research project has been supported by the National
Science Centre (Poland) through the grant No.~2022/45/B/ST3/02826.
\end{acknowledgements}

\bibliography{bibliography}
\end{document}